# AI Literacy and LLM Engagement in Higher Education: A Cross-National Quantitative Study

**Shahin Hossain**
University of Maryland Baltimore County
Baltimore, MD, United States
Email: shahinh1@umbc.edu

**Shapla Khanam**
Washington University of Science and Technology, VA, United States
Email: s.khanam@research.wust.edu

**Samaa Haniya**
Pepperdine University, Graduate School of Education and Psychology
Los Angeles, CA, United States
Email: samaa.haniya@pepperdine.edu

**Nesma Ragab Nasr**
Northern Arizona University
Flagstaff, AZ, United States
Email: nrn55@nau.edu

## Abstract

This study presents a cross-national quantitative analysis of how university students in the United States and Bangladesh interact with Large Language Models (LLMs). Based on an online survey of 318 students, results show that LLMs enhance access to information, improve writing, and boost academic performance. However, concerns about overreliance, ethical risks, and critical thinking persist. Guided by the AI Literacy Framework, Expectancy-Value Theory, and Biggs' 3P Model, the study finds that motivational beliefs and technical competencies shape LLM engagement. Significant correlations were found between LLM use and perceived literacy benefits ($r = .59$, $p < .001$) and optimism ($r = .41$, $p < .001$). ANOVA results showed more frequent use among U.S. students ($F = 7.92$, $p = .005$) and STEM majors ($F = 18.11$, $p < .001$). Findings support the development of ethical, inclusive, and pedagogically sound frameworks for integrating LLMs in higher education.

# Introduction

Generative artificial intelligence (GenAI) systems, particularly Large Language Models (LLMs) engineered to produce human-like texts, have made rapid advances since the public launch of OpenAI's ChatGPT in November 2022 and subsequent models such as GPT-4o, Google's Gemini Advanced, Anthropic's Claude, Meta's LLaMA, and DeepSeek. College students are increasingly using these tools for research support, content creation, and writing assistance, not only to streamline tasks but to reconceptualize access to information, idea expression, and literacy practices (Chen, 2023; Dai et al., 2023; Sharma, 2025).

Despite the increasing adoption of LLMs, there remains a lack of cross-national empirical research that examine how university students in developed and developing countries perceive and use these tools, particularly in terms of their motivational triggers, literacy learning, and ethical awareness (Ennion & McLellan, 2025; Cotton et al., 2024; Chiu, 2024). (Ennion & McLellan, 2025; Cotton et al., 2024; Chiu, 2024). Although the literature shows LLMs' promise for personalized learning support and overcome writers' block, persistent concerns about accuracy, algorithmic bias, misuse, and academic integrity underscore the need for deeper investigation (Chan & Hu, 2023; Giray et al., 2025; Mogavi et al., 2024).

The present study fills this gap through a cross-national quantitative survey of 318 students at two universities (United States, n = 171; Bangladesh, n = 147). We examine students' AI literacy, familiarity with LLMs, perceived utility, motivational beliefs, and self-reported impacts on academic performance and literacy practices. Our guiding question is: How do AI literacy, motivational beliefs, and educational context influence university students' engagement with Large Language Models, and what are the perceived academic and ethical outcomes of this engagement across cross-national and disciplinary settings.

To address this question, we integrate three complementary frameworks. The AI Literacy Framework (Allen & Kendeou, 2024; Long & Magerko, 2020) accounts for students' cognitive, technical, and ethical competencies in using LLMs. Expectancy-Value Theory (Wigfield & Eccles, 2000) informs our assessment of motivational dimensions such as perceived usefulness and outcome expectations. Finally, Biggs' 3P Model (Presage, Process, Product; Biggs et al., 2022) offers a lens for understanding how background factors and learning strategies mediate the use of LLMs and their outcomes.

By bringing together robust statistical analyses with cross-country perspectives, this work establishes evidence-based insights to inform curricula for educating in AI literacy, support equity-informed interventions, and discipline-specific pedagogy in a manner that facilitates effective and responsible integration of LLMs in higher education.

**Literature Review**

The adoption of LLMs in higher education has exponentially expanded because of their capabilities to enhance personalized learning, foster innovative literacy practices, and elevate academic productivity. Recent empirical research finds that college students increasingly value LLMs for their ability to support academic writing, streamline research processes, and facilitate the completion of college assignments (Al-Abri, 2025; Chakrabarty et al., 2024; Chu et al., 2025; Meyer et al., 2024). Additional conceptual studies, such as a systematic review of LLMs in higher education conducted by Wu et al. (2025) and a meta-analysis by Liu et al. (2025), confirm these findings and indicate strong correlations between formalized LLM interaction and improvements in learners' self-efficacy, creative potential, and academic productivity. Thus, LLMs are increasingly transforming traditional pedagogic frameworks through new understandings of literacy use and learning processes, while altering teacher-student dynamics and learning outcomes.

In spite of these promising developments, deep-seated anxieties prevail. Issues surrounding academic integrity, content accuracy, AI plagiarism, algorithmic biases, and potential detrimental impacts on critical thinking and cognitive development have drawn increased scholarly attention. MIT Lab's recent study conducted by Kosmyna et al. (2025), for example, finds that while LLMs like ChatGPT offer convenience, their application can diminish deep cognitive engagement and creative thinking, especially when used habitually. The researchers raise caution about learners' independent, autonomous, and critical thinking skills being worn down when learners continuously use AI tools for composition writing. Aside from that, empirical studies (Azoulay et al., 2025; Chan & Hu, 2023; Ong et al., 2024; Vetter et al., 2024) point out students' concerns regarding heavy reliance, privacy, ethical accountability, and academic integrity, despite recognition of LLMs' educational advantages that LLMs make the academic tasks easier. Current studies reinforce these apprehensions, notably Acut et al.'s (2025) book chapter, which emphasizes growing unease among educators regarding LLM use because of their potential to reduce student interaction in learning environments.

Within this evolving policy and research discourse, scholars increasingly stress the critical role of AI literacy as foundational for effective student engagement with LLM technologies. Allen and Kendeou (2024) argue for comprehensive educational interventions that foster digital competencies, ethical awareness, and operational knowledge of LLM models to mitigate biases, protect data privacy, and uphold intellectual integrity. Similarly, Dahlkemper et al. (2023) and Raman (2025) advocate for explicit integration of structured AI literacy frameworks in higher education curricula to address perceived educational and ethical problems in a structured approach. Recent scholarship by Chiu (2024) and Creely et al., (2025) provides strong evidence to warrant an urgent emphasis on interventions in AI literacy, directly addressing learners' cognitive, ethical, and technical skills to enable their practical and responsible application of AI in learning environments.

Yet, despite growing insights, a prominent gap remains concerning cross-national comparative analyses, particularly studies that examine underrepresented educational contexts. Current research disproportionately emphasizes Western-centric perspectives, with limited exploration of comparative cultural and institutional nuances that influence LLM adoption and usage. This oversight impedes the development of inclusive, context-sensitive educational policies and practices, especially in regions with markedly different academic traditions and technological infrastructure, such as the United States and Bangladesh.

To address this critical research gap, this study implements a rigorous quantitative examination of how university students from the United States and Bangladesh engage with LLM technologies. By analyzing students' familiarity, LLM-oriented motivational orientations, the perceived influences of LLM literacy on their lives, and their ethical considerations regarding the use of LLMs, this work generates significant cross-country comparative evidence. Through this process, this research enriches academic understanding and informs higher education policy analysts and teachers in designing culturally competent, inclusive AI-integrated curricula. This study thus makes a significant contribution to evidence-based practices designed to leverage the educational potential of LLMs responsibly, ethically, and equitably within diverse international contexts.

**Theoretical Framework**

This study is grounded in three complementary theoretical frameworks: the AI Literacy Framework, Expectancy-Value Theory, and Biggs' 3P Model. Together, these frameworks offer a comprehensive, multidimensional lens to understand not only students' preparedness for engaging with LLMs but also their motivational orientations and patterns of academic use. Each framework addresses a unique facet of LLM engagement: AI Literacy captures students' technical and ethical competencies, Expectancy-Value Theory explains their motivational drivers, and Biggs' 3P Model situates their engagement in a dynamic learning process. Their integration strengthens both the explanatory power and practical implications of this study.

**AI Literacy Framework**

At the core of this inquiry lies the AI Literacy Framework (Allen & Kendeou, 2024; Long & Magerko, 2020), which defines AI literacy as the ability to critically understand, evaluate, and utilize AI technologies in informed and ethical ways. This framework encompasses three domains: knowledge (awareness of AI systems, limitations, and ethical dimensions), skills (ability to use and assess AI outputs critically), and dispositions (ethical orientation, adaptability, and reflective judgment). Applied here, the framework enables an analysis of how students' AI literacy levels shape their interactions with LLMs ranging from uncritical reliance to strategic, reflective engagement.

**Figure 1.** *AI Literacy Framework*

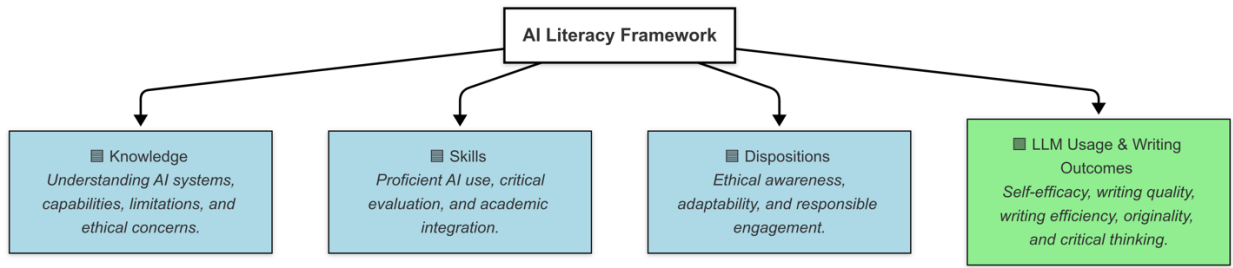

Figure 1 illustrates how knowledge, skills, and dispositions work in tandem to mediate effective, ethical use of LLMs in academic contexts. While the AI Literacy Framework offers crucial insight into students' readiness to engage responsibly with LLMs, it does not explain *why* students choose to use these tools. To capture motivational variation, we turn to Expectancy-Value Theory.

**Expectancy-Value Theory**

Complementing AI literacy, Expectancy-Value Theory (Wigfield & Eccles, 2000) provides insights into students' motivational factors influencing their engagement with LLMs. Expectancy-Value Theory proposes that students' willingness to utilize LLM tools is driven by their expectations for success (self-efficacy in using AI tools), perceived task value (practical and intrinsic benefits of AI tools), moderated by perceived costs (ethical concerns, risks of skill erosion, and institutional restrictions). In the context of LLM use, Expectancy-Value Theory helps explain individual differences in engagement by highlighting how students weigh anticipated benefits (e.g., improved writing, time-saving), confidence in using AI tools, and perceived drawbacks (e.g., ethical concerns, reduced skill development, institutional restrictions).

**Figure 2.** *Expectancy Value Theory and LLMs*

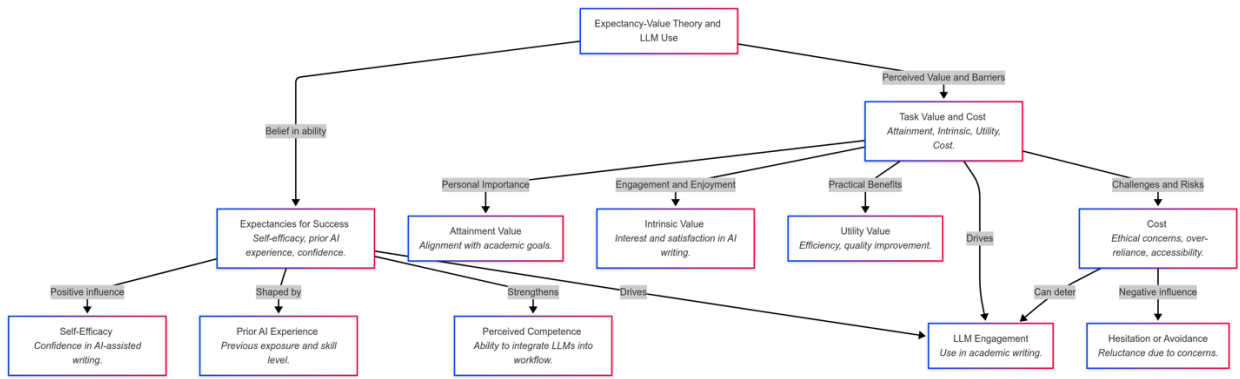

Figure 2 visualizes how expectancies, value appraisals, and perceived costs converge to influence students' motivation to engage with LLMs. Expectancy-Value Theory complements the AI Literacy Framework by offering a robust account of why students choose to engage with LLMs at different levels. However, it does not address how these students actually operationalize their engagement. To that end, Biggs' 3P Model adds a critical process-oriented perspective.

**John Biggs' 3P Model**

To further contextualize student engagement with LLMs, Biggs' 3P Model (Biggs, 1999; Biggs & Tang, 2022) frames learning as a dynamic process consisting of Presage (prior knowledge, AI literacy, and motivational beliefs), Process (strategies for engaging with AI tools), and Product (outcomes such as writing quality and academic integrity). This model emphasizes the importance of students' initial competencies and attitudes toward AI (Presage), their strategic and critical use of LLMs during the writing process (Process), and how these factors impact their academic outcomes (Product). Applying the 3P model allows the study to comprehensively examine how students' initial perceptions and capabilities influence their use of LLMs and subsequent academic performance.

**Figure 3.** *LLMs through John Bigg's 3P Model*

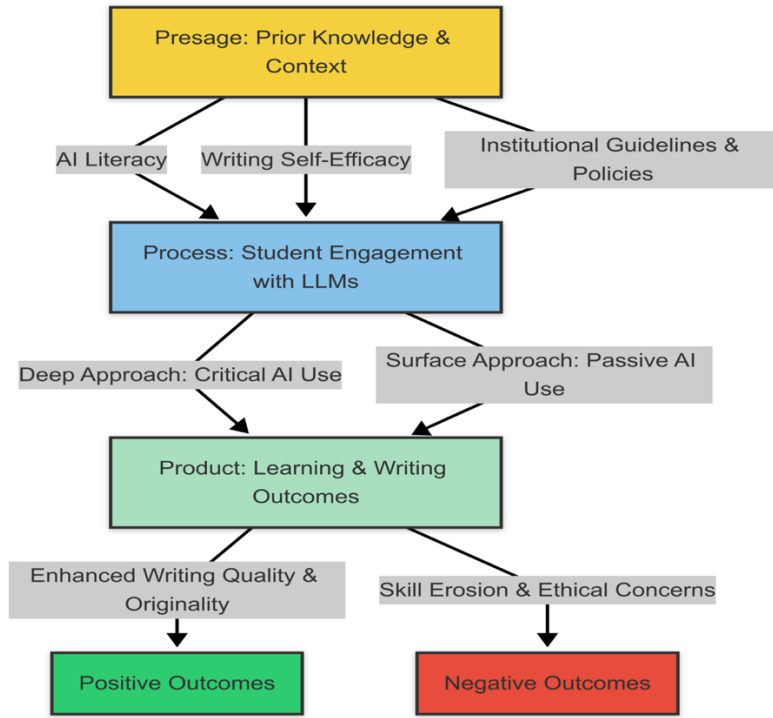

Figure 3 shows how learner characteristics interact with engagement strategies to produce varying academic outcomes. Biggs' model is indispensable for understanding how students engage with LLMs in authentic academic settings something neither the AI Literacy Framework nor Expectancy-Value Theory fully accounts for.

**Integrated Conceptual Framework**

These three integrated frameworks create a comprehensive conceptual lens through which this study examines students' use of LLMs. The AI Literacy Framework outlines core competencies for the responsible use of AI. Expectancy-Value Theory reveals the motivational forces behind engagement levels, and Biggs' 3P Model provides a robust approach for understanding the interaction among initial conditions, engagement processes, and academic outcomes. Collectively, these frameworks enable a multidimensional analysis of how students strategically leverage LLMs in diverse educational and cultural contexts, offering critical insights for educators and policymakers aiming to successfully integrate AI tools into higher education in an effective and responsible way.

## Research Methodology

**Research Design**

This study employed a cross-national, explanatory quantitative research design to investigate university students' perceptions, motivations, and engagement patterns with LLMs in higher education. The survey design was selected due to its application in capturing self-reported attitudes, competency levels, and behaviors across large, diverse student populations. The design was informed by a triangulated theoretical framework, comprising the AI Literacy Framework, Expectancy-Value Theory, and Biggs' 3P Model, that guided the selection and operationalization of key variables. The independent variables included AI literacy dimensions and motivational beliefs, while the dependent variables included LLM usage frequency and perceived academic outcomes. The design was non-experimental and correlational in nature, aiming to test associations and group differences rather than causal relationships. Figure 4 illustrates the structured workflow of this research.

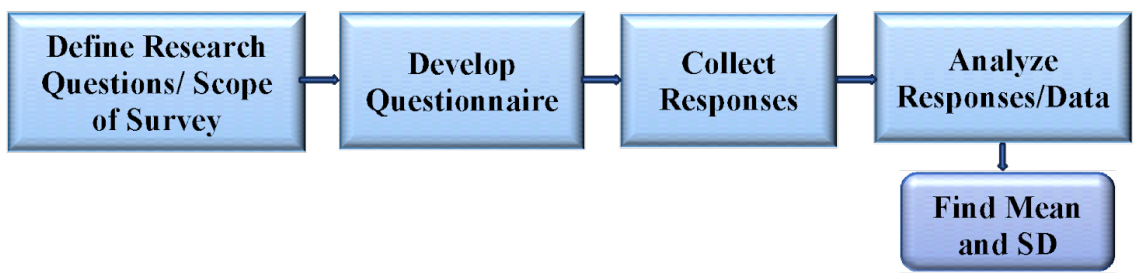

**Figure 4**: *Methodology*

**Population and Sampling**

Participants consisted of undergraduate and graduate students in two research-intensive universities in Bangladesh and the United States. We employed a purposive sampling procedure to ensure diversity in academic discipline (non-STEM vs. STEM), degree level (undergraduate vs. graduate), gender, and additional demographic characteristics. Participants were eligible if they were enrolled in degree seeking programs that permitted the use of AI tools for academic support. We obtained a total of 318 complete and valid responses, comprising 171 from the United States and 147 from Bangladesh. We conducted a power analysis and found that at least 146 valid responses were needed to meet conventional thresholds for detecting medium effect sizes in correlational and group comparison analyses.

**Instrumentation and Data Collection**

Data were collected through an online survey administered between April 2024 and April 2025. The survey instrument comprised 21 closed questions that were divided into five primary constructs: (1) AI literacy (knowledge, skills, and disposition); (2) familiarity and frequency of LLM use; (3) motivational beliefs (expectancies, task value, and perceived cost); (4) perceived academic impact; and (5) ethical concerns. All items, except the familiarity with AI, used a five-point Likert scale ranging from "strongly disagree" to "strongly agree."

Survey items were adapted from previously validated instruments in education technology, ethical applications of AI, and student motivation (e.g., Bråten et al., 2023; Chan & Hu, 2023; Chiu et al., 2024). A panel of three subject matter experts, two from the U.S. and one from Bangladesh, reviewed the instrument to ensure cross-cultural relevance, content validity, and construct alignment. Pilot testing was conducted with 43 respondents, incorporating minor revisions based on psychometric feedback and researchers' insights. Instrument reliability was determined through internal consistency testing.

**Data Analysis**

The scale of familiarity with AI, initially scored on a 10-point continuum, was rescaled to align with the other constructs. The statistical analysis was conducted using Python 3.11.0, with libraries such as Pandas for data structuring, matplotlib and seaborn for visualization, and scipy.stats and statsmodels for inferential testing.

Descriptive statistics (means, standard deviations, frequencies) were used to characterize the sample and main variables. Cronbach's alpha coefficients for each construct were calculated to ensure internal consistency; all exceeded the .70 reliability threshold (AI Literacy $\alpha$ = .77; Motivational Beliefs $\alpha$ = .81). Bivariate Pearson correlation analyses were conducted to examine relationships among AI literacy, motivation, LLM use, and perceived outcomes. Independent samples t-tests and one-way ANOVAs were performed to compare groups, particularly to identify differences based on nationality, academic field of study, and degree level. Post-hoc analyses (Tukey's HSD) and Levene's tests for homogeneity of variance were conducted to validate findings. A significance level of $p < .05$ was maintained across all tests.

**Ethical Considerations**

Ethical approval was obtained from each respective participating university's Institutional Review Board (IRB) prior to collecting the data. Participants received a digital informed consent document that specified the study's purpose, the voluntariness of participation, the risks and benefits associated with participation, and plans for maintaining confidentiality and anonymity. Participation was entirely voluntary, and no incentives were offered. However, 20 participants were each given $25 based on a random lottery draw. Data was anonymously collected with no

personally identifiable information being saved. Data was stored in securely controlled servers, and the dataset was de-identified before analysis. The study was based on principles of beneficence, non-maleficence, and respect for persons.

**Limitations and Bias**

While the study employed rigorous design and analysis procedures, various limitations need to be acknowledged. Purposive sampling limits generalizability to other institutions. Data are self-reported and susceptible to social desirability and recall inaccuracies. Cultural differences in the interpretation of individual survey questions can impact construct validity, even with expert review and pilot testing. A cross-sectional design also precludes causal inference. These limitations ultimately stem from the cross-sectional design, which precludes causal inference. These limitations are addressed through transparency in design and interpretation and offer avenues for future longitudinal or mixed-methods research.

## RESULTS

**Demographic Characteristics**

Table 1 summarizes the demographic characteristics of the 318 university students who participated in the study. Participants were nearly evenly distributed by country, with 54% from the United States and 46% from Bangladesh. In terms of gender identity, 51% identified as male, 45% as female, and 4% as non-binary. Age distribution was diverse, with 43% aged 22–25 years, and 32% aged 26 years or older. The sample primarily comprised Asian students (61%), followed by White (23%), Black/African American (7%), Hispanic/Latine (6%), and Mixed Race (4%). Academic level representation ranged from freshman to graduate students, with graduate students making up 34% of the sample. The disciplinary breakdown showed a balanced distribution between STEM (52%) and non-STEM (48%) majors, and language backgrounds were diverse, with 62% of students reporting a primary language other than English. This diversity supported meaningful subgroup analyses.

**Table 1.** *Demographic Characteristics of Respondents (N = 318)*

| Category | Subgroup | n | % |
|---|---|---|---|
| Country | United States | 171 | 54 |
| Country | Bangladesh | 147 | 46 |
| Gender | Man | 163 | 51 |
| Gender | Woman | 143 | 45 |
| Gender | Non-binary | 12 | 4 |
| Age | 18–21 years | 80 | 25 |
| Age | 22–25 years | 136 | 43 |
| Age | 26 years or older | 102 | 32 |
| Race/Ethnicity | Asian | 171 | 61 |
| Race/Ethnicity | Black/African American | 21 | 7 |
| Race/Ethnicity | Hispanic/Latine | 19 | 6 |
| Race/Ethnicity | Mixed Race | 11 | 4 |
| Race/Ethnicity | White | 73 | 23 |
| Academic Level | Freshman (1st year) | 11 | 4 |
| Academic Level | Sophomore (2nd year) | 83 | 26 |
| Academic Level | Junior (3rd year) | 66 | 21 |
| Academic Level | Senior (4th year) | 49 | 15 |
| Academic Level | Graduate | 109 | 34 |
| Major | STEM | 166 | 52 |
| Major | Non-STEM | 152 | 48 |
| Language Spoken | English | 121 | 38 |
| Language Spoken | Other than English | 197 | 62 |

**Descriptive Statistics**

Table 2 and Figure 5 present descriptive statistics for key variables. The mean frequency of LLM use (M = 3.46, SD = 1.00) indicated moderate engagement among students. Students expressed high optimism about the academic potential of LLMs (M = 3.68, SD = 0.47) and reported a moderately high perceived impact on their literacy skills (M = 2.94, SD = 0.65). Despite this, self-reported AI literacy was notably low (M = 1.86, SD = 1.64), highlighting limited formal training. AI familiarity scores (M = 3.72, SD = 0.93) indicated moderate familiarity, suggesting room for further educational enhancement.

**Table 2.** *Descriptive Statistics of Key Variables*

| Variable | Mean | SD |
|---|---|---|
| AI Literacy | 1.86 | 1.64 |
| AI Familiarity | 3.72 | 0.93 |
| LLM Use Frequency | 3.46 | 1.0 |
| Perceived Literacy Impact | 2.94 | 0.65 |
| Grade Impact | 4.08 | 0.49 |
| Ethical Concern | 3.17 | 0.89 |
| Optimism | 3.68 | 0.47 |

**Figure 5.** *Mean and Standard Deviation of Key LLM Variables*

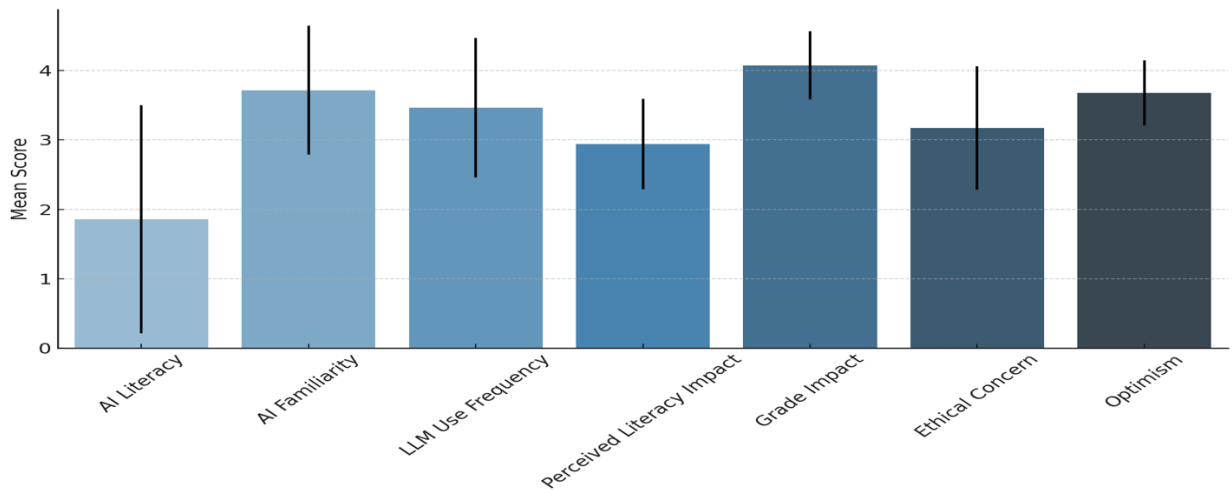

**Correlational Analysis**

Pearson correlation coefficients among the key study variables are depicted in Figure 6. Significant correlations emerged between LLM use frequency and perceived literacy impact ($r = .59, p < .001$), optimism ($r = .41, p < .001$), and AI familiarity ($r = .40, p < .001$). These correlations reflect medium to large effects and indicate that frequent users of LLMs perceive stronger literacy benefits and exhibit higher optimism toward AI integration. Ethical concern exhibited a weak negative correlation with AI literacy (r = –.18, p < .05), suggesting that increased literacy slightly decreases ethical apprehension.

**Figure 6.** *Correlation Matrix of Key Constructs*

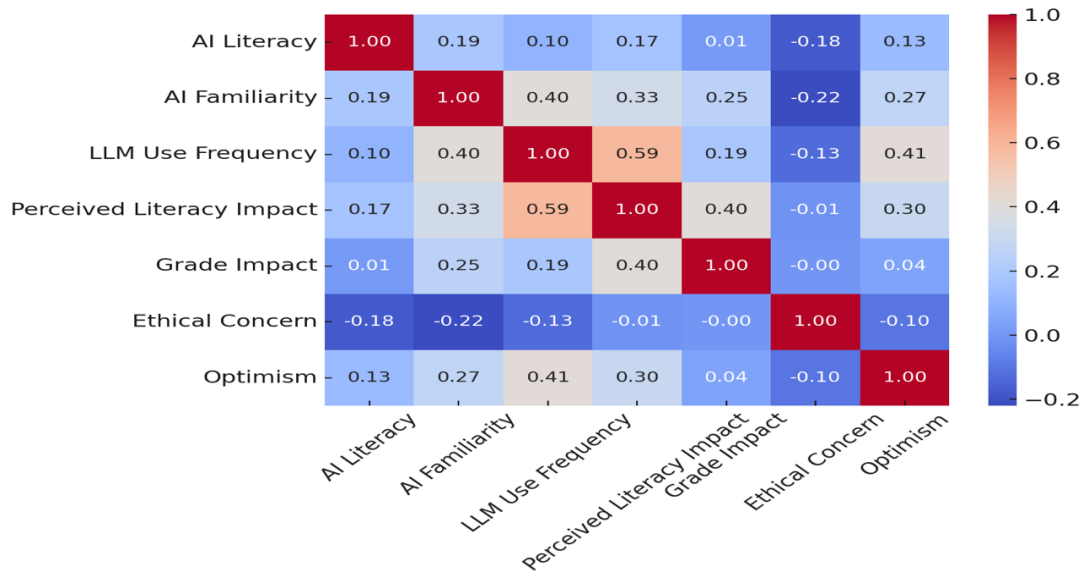

**Group Comparisons**

Group differences were analyzed using independent samples t-tests and ANOVAs, with results summarized in Table 3. Male students (M = 3.67, SD = 0.91) reported significantly higher LLM use frequency than female students (M = 3.24, SD = 1.03), t(298) = 2.66, p = .008, Cohen's d = 0.44 (medium effect). U.S. students (M = 3.61, SD = 0.95) used LLMs more frequently compared to Bangladeshi students (M = 3.26, SD = 1.03), F(1,316) = 7.92, p = .005, $\eta^2$ = .025 (small-to-medium effect). STEM students (M = 3.75, SD = 0.85) also reported significantly higher usage than non-STEM students (M = 3.14, SD = 1.04), F(1,316) = 18.11, p < .001, $\eta^2$ = .054 (medium effect).

**Table 3.** *Inferential Results Summary*

| Comparison Type | Variable | Test (t/F) | p-value | Significant? | Summary |
|---|---|---|---|---|---|
| t-test (Gender) | LLM Use Frequency | t(298) = 2.66 | .008 | Yes | Males use LLMs more frequently |
| ANOVA (Country) | LLM Use Frequency | F(1,316) = 7.92 | .005 | Yes | U.S. students use LLMs more frequently |
| ANOVA (Major) | LLM Use Frequency | F(1,316) = 18.11 | < .001 | Yes | STEM > non-STEM |
| t-test (Country) | Perceived Literacy Impact | t = −4.59 | < .001 | Yes | Bangladesh > U.S. |
| t-test (Country) | Optimism | t = 3.68 | < .001 | Yes | U.S. > Bangladesh |
| t-test (Country) | AI Literacy | t = 1.81 | .070 | No | Trend (U.S. > BD) |
| t-test (Country) | AI Familiarity | t = 0.89 | .372 | No | No difference |
| t-test (Country) | Grade Impact | t = −0.88 | .379 | No | No difference |
| t-test (Country) | Ethical Concern | t = −1.47 | .143 | No | No difference |

Additional country-level comparisons highlighted that Bangladeshi students (M = 3.12, SD = 0.56) perceived significantly greater literacy benefits from LLMs compared to U.S. students (M = 2.78, SD = 0.68), t(316) = –4.59, p < .001, Cohen's d = 0.56 (medium effect). Conversely, U.S. students (M = 3.79, SD = 0.44) expressed significantly higher optimism regarding LLM integration compared to Bangladeshi students (M = 3.54, SD = 0.49), t(316) = 3.68, p < .001, Cohen's d = 0.54 (medium effect). No significant differences emerged between countries regarding AI literacy, AI familiarity, grade impact, or ethical concerns.

Figures 7 and 8 visually illustrate these significant group differences, further clarifying how LLM engagement, literacy impact, and optimism vary across national and disciplinary contexts.

**Figure 7.** *Cross-National Comparison of Mean Scores (U.S. vs. Bangladesh)*

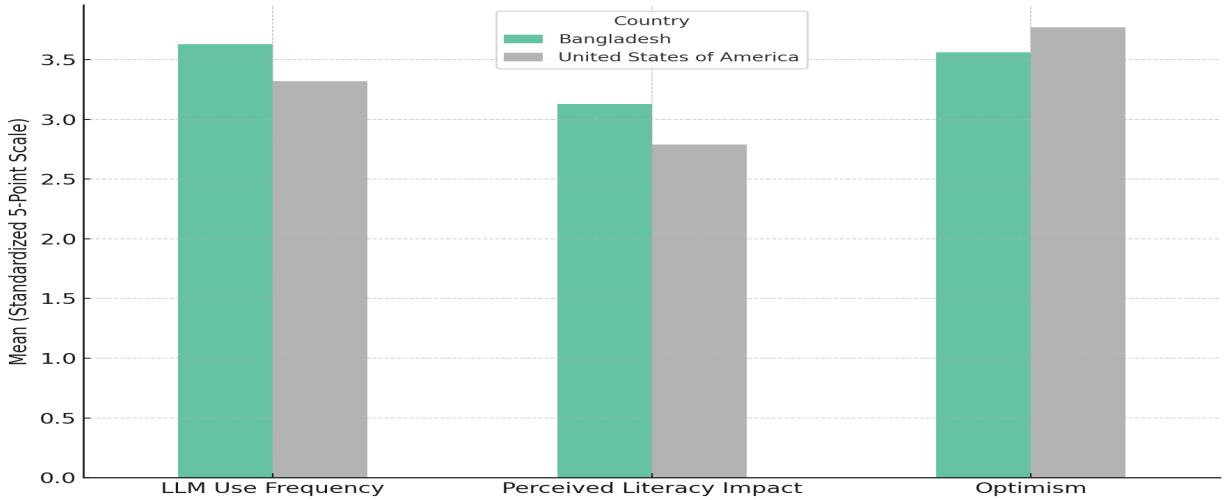

**Figure 8.** *STEM vs. Non-STEM Comparison of Mean Scores*

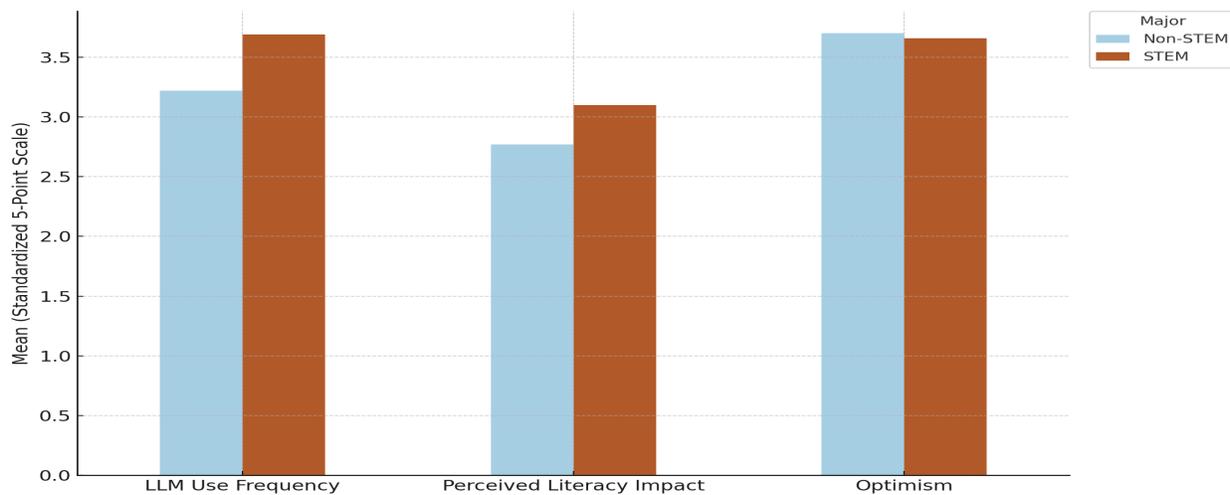

Collectively, these descriptive, correlational, and comparative findings provide a comprehensive empirical foundation. They underscore how demographic and disciplinary factors shape students' engagement with LLMs, highlighting key motivational, ethical, and literacy-related trends. These results lay a robust groundwork for the discussion of implications, educational interventions, and policy recommendations in subsequent sections.

## DISCUSSION

This research provides robust empirical insight into how graduate and undergraduate students from America and Bangladesh engage with LLMs in an academic environment. Using the AI Literacy Framework, Expectancy-Value Theory, and Biggs' 3P Model of learning, this research identifies key relationships between students' AI literacy, motivational beliefs, literacy practice, and ethical consideration. The study has considerable policy relevance for institutions, curriculum development in universities, and responsible integration of LLMs in higher education.

In spite of widespread adoption of LLM tools such as Gemini and ChatGPT, formal AI literacy among learners remains significantly deficient. In line with global research (Allen & Kendeou, 2024; Chiu, 2024), this finding indicates a critical gap in foundational AI education demonstrated by low scores in formal training ($M = 1.86$, $SD = 1.64$). Low levels of AI literacy can exacerbate ethical concerns ($M = 3.17$) because students may not be aware of ethical issues such as authorship, algorithmic bias, and overreliance on AI tools for academic writing. The weak negative correlation between ethical issues and AI literacy ($r = -.18$) translates to most learners being ignorant of complexity and seriousness of the ethical problems in AI. These necessitate institutions to make concerted efforts toward holistic education in AI literacy with particular instruction in ethical thinking, critical evaluation of AI outputs, and responsible AI use.

The motivational findings closely align with Expectancy-Value Theory (Wigfield & Eccles, 2000). Significant correlations between LLM usage frequency and perceived literacy impact ($r = .59$, $p < .001$) and optimism ($r = .41$, $p < .001$) underscore perceived utility as a fundamental driver of student engagement with LLMs. Notably, cross-national differences emerged clearly, with U.S. students demonstrating higher optimism and Bangladeshi students reporting greater perceived literacy benefits. These distinctions suggest that institutional policies must address contextual motivational drivers, enhancing transparency, highlighting practical benefits, and establishing supportive environments to foster effective student engagement with AI tools.

The application of Biggs' 3P Model also explained how presage factors (e.g., familiarity with AI, disciplinary orientation) considerably affected students' LLM utilization and perceived results. Students majoring in STEM disciplines have greater exposure to AI tools and reported higher LLM use than their non-STEM counterparts ($\eta^2 = .054$, medium effect).

On the other hand, gender differences in use also emerged significantly, with higher LLM utilization found among male students (Cohen's $d = 0.44$, medium effect). These results show the necessity for customized learning strategies that cater specifically to diverse student profiles and learning contexts.

Based on these empirical insights, this study proposes four targeted policy and practice recommendations. First, higher education institutions should institutionalize comprehensive AI literacy curricula, incorporating robust, ethically informed training across all academic disciplines to systematically address current gaps. Second, to promote equity and inclusive access, institutions should develop culturally and linguistically inclusive programs, ensuring that all students have equitable opportunities for AI-related training and technology access. Third, discipline-specific pedagogical strategies should be implemented, creating tailored approaches to enhance the relevance and adoption of LLM tools, particularly among students in non-STEM disciplines. Finally, institutions are encouraged to develop culturally adaptable AI policy frameworks that recognize and integrate context-specific differences in student optimism and perceptions of literacy benefits across diverse national and institutional settings.

Several limitations of this study merit consideration. While the cross-national design provides valuable comparative insights, it does not fully capture institutional variability within each country or changes over time. The reliance on self-reported data introduces potential biases such as social desirability and subjective interpretation. Future research should incorporate mixed-method approaches and longitudinal studies to capture evolving attitudes and practices regarding LLM use comprehensively.

As generative AI increasingly reshapes higher education, this study offers timely, practical insights into student engagement with LLM technologies. Emphasizing robust AI literacy, motivational alignment, and culturally responsive policy-making, these findings support institutions in effectively integrating AI tools, enhancing ethical practice, and maximizing positive educational outcomes.

## CONCLUSION

This study investigated university students' perceptions, motivations, and usage patterns of LLMs, examining their implications for literacy practices and academic outcomes across diverse institutional contexts in Bangladesh and the United States. Employing an integrated theoretical framework comprising the AI Literacy Framework (Allen & Kendeou, 2024; Long & Magerko, 2020), Expectancy-Value Theory (Wigfield & Eccles, 2000), and Biggs' 3P Model (Biggs, 1999), this research provided a comprehensive analysis of how foundational AI knowledge, motivational orientations, and learning experiences collectively shape students' interactions with LLMs.

Quantitative findings underscored a significant gap in formal AI literacy among students, as evidenced by low average scores in AI training ($M = 1.86$), and highlighted moderate ethical concerns ($M = 3.17$). Strong positive correlations (e.g., $r = .59$ between LLM use frequency and perceived literacy benefits; $r = .41$ between usage and optimism) revealed that students' perceived usefulness and optimism substantially drive their engagement. Furthermore, cross-national and disciplinary differences emerged clearly, underscoring the need for culturally sensitive and discipline-specific pedagogical strategies.

The results highlight several critical implications for policy and practice in higher education. Institutions must prioritize comprehensive AI literacy curricula that emphasize ethical, responsible, and critical engagement with AI technologies. Equity-driven initiatives should ensure inclusive and accessible AI education across linguistic and cultural boundaries, while discipline-specific approaches should facilitate the effective integration of LLMs into diverse academic contexts. Additionally, culturally adaptable AI policy frameworks are crucial for addressing the varying levels of student optimism, literacy impacts, and ethical concerns across international contexts.

Future research should adopt longitudinal designs to better understand the evolving interactions between students and LLMs and investigate institutional variations within and across countries. Expanding the scope of analysis to incorporate qualitative dimensions could further enrich insights into how students' ethical considerations and learning strategies develop over time. By addressing these avenues, subsequent studies will contribute valuable depth to our understanding of how to responsibly integrate LLMs into educational environments.

This research advances the scholarship on AI in education by demonstrating the complex interplay among AI literacy, student motivation, and institutional context. Its findings inform evidence-based strategies for higher education institutions, promoting the ethical integration of AI that not only enhances students' literacy and learning outcomes but also safeguards critical thinking and academic integrity in the era of GenAI.

# REFERENCES

Acut, D. P., Gamusa, E. V., Pernaa, J., Yuenyong, C., Pantaleon, A. T., Espina, R. C., Sim, M. J., & Garcia, M. B. (2025). AI Shaming Among Teacher Education Students: A Reflection on Acceptance and Identity in the Age of Generative Tools. In M. Garcia, J. Rosak-Szyrocka, & A. Bozkurt (Eds.), *Pitfalls of AI Integration in Education: Skill Obsolescence, Misuse, and Bias* (pp. 97-122). IGI Global Scientific Publishing. https://doi.org/10.4018/979-8-3373-0122-8.ch005

Al-Abri, A. (2025). Exploring ChatGPT as a virtual tutor: A multi-dimensional analysis of large language models in academic support. *Education and Information Technologies*. https://doi.org/10.1007/s10639-025-13484-x

Allen, L. K., & Kendeou, P. (2024). ED-AI Lit: An Interdisciplinary framework for AI literacy in education. *Policy Insights from the Behavioral and Brain Sciences*, *11*(1), 3-10. https://doi.org/10.1177/23727322231220339

Azoulay, R., Hirst, T., & Reches, S. (2025). Large Language Models in Computer Science Classrooms: Ethical Challenges and Strategic Solutions. *Applied Sciences (2076-3417)*, *15*(4). DOI:10.3390/app15041793

Biggs, J. (1999). What the student does: Teaching for enhanced learning. *Higher Education Research & Development, 18*(1), 57-75. https://doi.org/10.1080/0729436990180105

Biggs, J., Tang, C., & Kennedy, G. (2022). *Teaching for quality learning at university* (5th ed.). McGraw-hill education (UK). ISBN-13: 9780335250820

Bråten, I., Haverkamp, Y. E., Latini, N., & Strømsø, H. I. (2023). Measuring multiple-source based academic writing self-efficacy. *Frontiers in Psychology*, *14*, 1-12. https://doi.org/10.3389/fpsyg.2023.1212567

Chakrabarty, T., Padmakumar, V., Brahman, F., & Muresan, S. (2024). Creativity Support in the Age of Large Language Models: An Empirical Study Involving Professional Writers. In *Proceedings of the 16th Conference on Creativity & Cognition* (pp. 132-155). https://doi.org/10.1145/3635636.3656201

Chan, C. K. Y., & Hu, W. (2023). Students' voices on generative AI: Perceptions, benefits, and challenges in higher education. *International Journal of Educational Technology in Higher Education*, *16*(2), 1-18. https://doi.org/10.1186/s41239-023-00411-8

Chen, S. Y. (2023). Generative AI, learning and new literacies. Journal of Educational Technology Development & Exchange, *16*(2). 1-19. https://doi.org/10.18785/jetde.1602.01

Chiu, T. K. (2025). Student AI Literacy and Competency. In *Empowering K-12 Education with AI*. Taylor & Francis. 30-54. DOI: 10.4324/ 9781003498377- 2

Chiu, T. K., Ahmad, Z., Ismailov, M., & Sanusi, I. T. (2024). What are artificial intelligence literacy and competency? A comprehensive framework to support them. *Computers and Education Open*, *6*, 100171. https://doi.org/10.1016/j.caeo.2024.100171

Chiu, T. K. (2024). Future research recommendations for transforming higher education with generative AI. *Computers and Education: Artificial Intelligence*, *6*, 100197. DOI:10.1016/j.caeai.2023.100197

Chu, Z., Wang, S., Xie, J., Zhu, T., Yan, Y., Ye, J., ... & Wen, Q. (2025). Llm agents for education: Advances and applications. *arXiv preprint arXiv:2503.11733*.

Cotton, D. R., Cotton, P. A., & Shipway, J. R. (2024). Chatting and cheating: Ensuring academic integrity in the era of ChatGPT. *Innovations in education and teaching international*, *61*(2), 228-239. https://doi.org/10.1080/14703297.2023.2190148

Creely, E., Barnes, M., Tour, E., Henderson, M., Waterhouse, P., Pena, M. A., & Patel, S. V. (2025). Exploring attitudes to generative AI in education for English as an additional language (EAL) adult learners. *ReCALL*, *37*(2), 174–190. doi:10.1017/S0958344024000314

Dahlkemper, M. N., Lahme, S. Z., & Klein, P. (2023). How do physics students evaluate artificial intelligence responses on comprehension questions? A study on the perceived scientific accuracy and linguistic quality of ChatGPT. *Physical Review Physics Education Research*, *19*(1), 010142. DOI: https://doi.org/10.1103/PhysRevPhysEducRes.19.010142

Dai, Y., Liu, A., & Lim, C. P. (2023). Reconceptualizing ChatGPT and generative AI as a student-driven innovation in higher education. *Procedia CIRP*, *119*, 84-90. https://doi.org/10.1016/j.procir.2023.05.002

Dempere, J., Modugu, K., & Hesham, A. (2023). The impact of ChatGPT on higher education. *Frontiers in Education,* *8*, 1-13. https://doi.org/10.3389/feduc.2023.1206936

Eccles, J. S., & Wigfield, A. (2020). From expectancy-value theory to situated expectancy-value theory: A developmental, social cognitive, and sociocultural perspective on motivation. *Contemporary educational psychology*, *61*, 101859. https://doi.org/10.1016/j.cedpsych.2020.101859

Ennion, M., & McLellan, R. (2025). Large Language Model Chatbots in education: Exploring literature insights on their impact and influence on learning behaviours. *Studies in Technology Enhanced Learning*, *4*(1). DOI:10.21428/8c225f6e.540b41b5

Giray, L., Sevnarayan, K., & Ranjbaran Madiseh, F. (2025). Beyond Policing: AI Writing Detection Tools, Trust, Academic Integrity, and Their Implications for College Writing. *Internet Reference Services Quarterly*, *29*(1), 83-116. https://doi.org/10.1080/10875301.2024.2437174

Kosmyna, N., Hauptmann, E., Yuan, Y. T., Situ, J., Liao, X. H., Beresnitzky, A. V., ... & Maes, P. (2025). Your Brain on ChatGPT: Accumulation of Cognitive Debt when Using an AI Assistant for Essay Writing Task. *arXiv preprint arXiv:2506.08872*.

Liu, X., Guo, B., He, W., & Hu, X. (2025). Effects of generative artificial intelligence on K-12 and higher education students' learning outcomes: A meta-analysis. *Journal of Educational Computing Research*. https://doi.org/10.1177/073563312513291

Long, D., & Magerko, B. (2020). What is AI literacy? Competencies and design considerations. *CHI '20: Proceedings of the 2020 CHI Conference on Human Factors in Computing Systems*, 1–16. https://doi.org/10.1145/3313831.3376727

Meyer, J., Jansen, T., Schiller, R., Liebenow, L. W., Steinbach, M., Horbach, A., & Fleckenstein, J. (2024). Using LLMs to bring evidence-based feedback into the classroom: AI-generated feedback increases secondary students' text revision, motivation, and positive emotions. *Computers and Education: Artificial Intelligence*, *6*, 100199. https://doi.org/10.1016/j.caeai.2023.100199

Mogavi, R. H., Deng, C., Kim, J. J., Zhou, P., Kwon, Y. D., Metwally, A. H. S., ... & Hui, P. (2024). ChatGPT in education: A blessing or a curse? A qualitative study exploring early adopters' utilization and perceptions. *Computers in Human Behavior: Artificial Humans*, *2*(1), 100027. https://doi.org/10.1016/j.chbah.2023.100027

Ong, J. C. L., Seng, B. J. J., Law, J. Z. F., Low, L. L., Kwa, A. L. H., Giacomini, K. M., & Ting, D. S. W. (2024). Artificial intelligence, ChatGPT, and other large language models for social determinants of health: Current state and future directions. *Cell Reports Medicine*, *5*(1). https://doi.org/10.1016/j.xcrm.2023.101356

Raman, R. (2025). Transparency in research: an analysis of ChatGPT usage acknowledgment by authors across disciplines and geographies. *Accountability in Research*, *32*(3), 277-298. https://doi.org/10.1080/08989621.2023.2273377

Sharma, S. (2025). Responding to Technology-Induced Transformations in Writing Education: Conceptualizing and Teaching the Literacies of Privacy, Originality, and Agency. *International Journal of Transformative Teaching and Learning in Higher Education*, *1*(1), 1-15. https://orcid.org/0000-0002-1456-8639

Vetter, M. A., Lucia, B., Jiang, J., & Othman, M. (2024). Towards a framework for local interrogation of AI ethics: A case study on text generators, academic integrity, and composing with ChatGPT. *Computers and composition*, *71*, 102831. https://doi.org/10.1016/j.compcom.2024.102831

Wigfield, A., & Eccles, J. S. (2000). Expectancy–value theory of achievement motivation. *Contemporary Educational Psychology*, *25*(1), 68–81. https://doi.org/10.1006/ceps.1999.1015

Wu, F., Dang, Y., & Li, M. (2025). A Systematic Review of Responses, Attitudes, and Utilization Behaviors on Generative AI for Teaching and Learning in Higher Education. *Behavioral Sciences*, *15*(4), 467. doi: 10.3390/bs15040467